# Three-dimensional visualization of lattice defects in β-$Ga_2O_3$ via synchrotron-radiation Borrmann-effect X-ray topo-tomography

Yongzhao Yao,[1,2,a)] Daiki Katsube,[2] Hirotaka Yamaguchi,[2] Shinya Yamaguchi,[3] Daiki Wakimoto,[3] Hironobu Miyamoto,[3] Yukari Ishikawa[2]

[1]Mie University, 1577 Kurimamachiya-cho, Tsu, Mie 514-8507, Japan
[2]Japan Fine Ceramics Center, 2-4-1 Mutsuno, Atsuta, Nagoya 456-8587, Japan
[3]Novel Crystal Technology, Inc., 2-3-1 Hirosedai, Sayama, Saitama 350-1328, Japan

**Abstract:** β-$Ga_2O_3$ is a promising material for next-generation power electronics; however, its performance is strongly affected by lattice defects such as dislocations. In this study, we demonstrate three-dimensional (3D) visualization of dislocations in β-$Ga_2O_3$ using synchrotron-radiation X-ray topo-tomography under a two-beam Borrmann-effect condition in transmission X-ray topography. By rotating the sample about the diffraction vector and acquiring a series of topo-tomographic images at different rotation angles, the evolution of dislocation contrast is captured, providing intuitive, depth-resolved visualization of dislocations. This method enables clear separation of dislocations in the epilayer and substrate in Schottky barrier diode structures, offering insight into dislocation propagation and their impact on epitaxial growth and device performance. This study represents the first demonstration of 3D dislocation reconstruction in β-$Ga_2O_3$.

[a)]Author to whom correspondence should be addressed. Electronic mail: yao@icsdf.mie-u.ac.jp.
ORCID:0000-0002-7746-4204

## I. INTRODUCTION

β-$Ga_2O_3$ has attracted significant attention as a next-generation material for power electronics owing to its outstanding physical properties, including a wide bandgap of approximately 4.8 eV, a high breakdown field strength of 7–8 MV/cm, and a superior Baliga figure of merit [1, 2, 3, 4]. Despite these advantages, the performance and reliability of β-$Ga_2O_3$-based devices are strongly degraded by the presence of lattice defects such as dislocations, stacking faults, domain boundaries, and pipe-like voids [5, 6, 7, 8]. A detailed understanding of these defects—particularly their formation, propagation, and interactions with interfaces during both bulk crystal growth and epitaxial growth—requires advanced characterization techniques capable of three-dimensional (3D) visualization. Conventional methods, including selective chemical etching [9, 10, 11, 12, 13, 14], X-ray topography (XRT) operated in either reflection [15, 16, 17, 18, 19, 20, 21] or transmission [22, 23, 24] configurations, phase-contrast microscopy [25], and transmission electron microscopy (TEM) [5, 26, 27, 28, 29, 30, 31], provide valuable insights but remain limited in their ability to deliver full 3D information. While techniques such as multi-photon excitation microscopy [32, 33, 34] and section topography [35, 36] have enabled 3D defect visualization in materials like GaN and SiC, they are not readily applicable to β-$Ga_2O_3$.

X-ray imaging is a powerful approach for probing defects in highly crystalline semiconductors, and we have previously demonstrated reflection XRT [16] and transmission XRT based on the Borrmann effect [24] (also known as the anomalous transmission) to observe dislocations in β-$Ga_2O_3$. Although variation of incident angle and diffraction vector allows partial depth sensitivity, complete 3D reconstruction remains challenging. A possible approach to obtain 3D structural information through X-ray imaging is X-ray computed tomography (CT) [37]. In CT, the sample is scanned stepwise over a range of angles to acquire a series of projection images, from which a 3D reconstruction is generated using algorithms based on the Radon transform [38]. However, because conventional CT relies on X-ray absorption contrast, it is inherently insensitive to lattice defects such as dislocations, making it unsuitable for their direct visualization. An extended form of CT that combines diffraction contrast with tomographic reconstruction is known as X-ray topo-tomography [39]. Unlike absorption-based CT, X-ray topo-tomography acquires two-dimensional distributions of the Bragg reflectivity within a crystal by precisely adjusting the orientation of the incident X-rays, the sample, and the detector to satisfy a specific Bragg diffraction condition. When the sample is rotated around the diffraction vector (***g***-vector) corresponding to the selected Bragg reflection—that is, the direction normal to the diffracting lattice planes—the local reflectivity remains effectively invariant with respect to rotation, provided that absorption due to sample thickness can be neglected. This condition is generally well satisfied in transmission XRT. As a result, reconstruction procedures analogous to those used in absorption-based CT can be applied, enabling the determination of a 3D map of the local Bragg reflectivity. Since lattice defects such as dislocations, stacking faults, domain boundaries, inclusions, and voids induce local variations in Bragg reflectivity, their spatial distribution and geometrical characteristics can be elucidated from the reconstructed 3D reflectivity within the crystal bulk. This technique has been successfully applied to visualize the 3D distribution of dislocations in materials such as diamond [39], silicon [40, 41, 42], and SiC [43]. These developments suggest that topo-tomographic approaches hold strong potential for achieving true 3D visualization of lattice defects in β-$Ga_2O_3$.

In this work, we applied X-ray topo-tomography under a two-beam Borrmann-effect condition in transmission XRT to visualize lattice defects in β-$Ga_2O_3$ for the first time.

Using this approach, we obtained animations of the 3D dislocation distribution in a rotating crystal, enabling intuitive and direct depth-resolved observation of dislocations in both β-$Ga_2O_3$ substrates and Schottky barrier diode (SBD) devices. This method provides insight into dislocation behavior and offers a useful approach for improving crystal quality and device performance.

## II. EXPERIMENTAL DETAILS

Two types of samples were investigated: a bare substrate and an SBD wafer. The bare substrate was a (001)-oriented β-$Ga_2O_3$ single crystal fabricated by the edge-defined film-fed growth (EFG) method. It was Sn-doped with a donor concentration of $8.4 \times 10^{18}$ $cm^{-3}$. The SBD wafer consisted of a Sn-doped (001)-oriented substrate, a 12 μm-thick epitaxial layer with n = $3.0 \times 10^{16}$ $cm^{-3}$ grown by hydride vapor phase epitaxy (HVPE), and electrode structures. Both samples had a thickness of 680–700 μm, corresponding to a $\mu t$ value of approximately 10, where $\mu$ is the linear absorption coefficient and $t$ is the sample thickness. Further details of the samples can be found in Refs. **15** and **22**.

X-ray topo-tomography was performed using a synchrotron-radiation monochromatic beam with a wavelength of λ = 1.24 Å at beamline BL-14B of the Photon Factory, High Energy Accelerator Research Organization (KEK), Japan. A photograph of the optical system is shown in Figure 1(a). The sample was mounted on a supporting frame attached to a six-axis goniometer (translation axes: x, y, z; rotation axes: ω, φ, χ), with a minimum ω step of 0.0002°. The (020) planes were used as reflecting planes, corresponding to a diffraction vector $\boldsymbol{g}$ = 020, in order to realize the Borrmann effect (Figure 1(c)). The Bragg angle ω for this reflection at the selected wavelength is 24°. Because the entrance and exit surfaces of the crystal were perpendicular to the (020) planes, the diffraction geometry corresponds to the symmetric Laue case. Under this condition, both the transmitted wave (o-wave) and diffracted wave (h-wave) are excited with comparable intensity (two-wave condition), which can be observed on a fluorescent screen in the absence of ambient light (Figure 1(b)). After establishing this condition, the fluorescent screen was removed to allow the forward-diffracted beam (o-wave) to enter the imaging system. The imaging system consists of a $Gd_3Al_2Ga_3O_{12}$ (GAGG:Ce) scintillator, a relay lens, and a CMOS camera [**44**]. The effective spatial resolution is approximately 2.2 μm/pixel. To acquire topo-tomography images, the sample was rotated stepwise about the diffraction vector $\boldsymbol{g}$ = 020, corresponding to the χ axis of the goniometer (Figure 1(a)). After each step of rotation, a fine adjustment of the ω angle was performed to re-establish the exact Bragg condition for the Borrmann effect. Topo-tomography simulations were carried out using in-house Python code.

## III. RESULTS AND DISCUSSION

A series of topo-tomographic images acquired at different χ angles serve as the basis for the 3D reconstruction of lattice defects. First, we describe the characteristic features of the XRT images and how the appearance of dislocations varies with χ. Topo-tomographic images acquired with χ rotated in 15° increments from 0° to 60° are shown in Figures 2(a)–2(e). These images are positive images, meaning that bright contrast corresponds to stronger X-ray exposure. The insets show photographs of the sample at each χ angle. In Borrmann-effect XRT images, dislocations appear as dark lines on a bright background, as they correspond to regions of displaced atomic arrangement, resulting in reduced anomalous transmission [**45**, **46**]. From Figure 2(a) to 2(e), the same dislocations appear in the images with different lengths and orientations depending on χ. Two sets of dislocations are selected as examples to illustrate these changes. The first set is the spindle-shaped dislocation marked by the two blue arrows in Figure 2(a). It becomes compressed along the [100] direction with increasing χ, and the inclination of its long axis gradually shifts toward the vertical direction in the image. The second set corresponds to dislocations No. 1–3, marked by the red arrows. In the absence of a clear intersection point as a reference, variations in their lengths are less apparent than in the first set; however, changes in their orientation are clearly evident, as indicated by the red arrows in Figure 2(e). The graphs below and above the XRT images show the geometrically simulated results for sets 1 and 2, respectively, as a function of χ, assuming that both the spindle-shaped dislocation and dislocations No. 1–3 lie on the (001) plane (see Supplementary Movies 1 and 2 for the simulation results). The wheel-like blue pattern serves as a visual guide and corresponds to a circle on the (001) plane with radial lines at 30° intervals. At χ = 0, dislocations No. 1–3 form angles of approximately −60°, 0°, and 30° with respect to the [100] direction, as estimated from this pattern. The experimental observations show good agreement with the simulations over the entire χ range until χ = 60° (Figure 2(e)), supporting the validity of the assumption that these dislocations lie on the (001) plane. It should be noted that the effective sample thickness for the transmitted o-wave increases with increasing χ, resulting in reduced transmission intensity and increased noise in the XRT images, particularly at χ = 60°. For the same reason, XRT images at χ > 60° were difficult to obtain.

Figure 3 shows static images of the 3D view, the camera view of the simulated results, and the experimentally acquired topo-tomographic images at χ = −20°. Readers are referred to Supplementary Movie 3 for an animation of these three datasets over the range of χ = −20° to +20°. In this experiment, the SBD wafer was observed. The χ increment was set to 0.5° per step, which is much finer than that used in Figure 2, in order to obtain a smooth animation and enable detailed identification of dislocation positions. The simulation was performed by rotating a model plate consisting of a three-layer mesh pattern in the same manner as rotating the real sample. The model plate thickness was set to 700 μm, identical to that of the sample. The three-layer mesh pattern was arranged in a 10 × 10 grid (blue grid lines) on planes parallel to sample surface at three z positions: 0, 350 μm, and 700 μm, corresponding to the front surface of the wafer (where the SBD electrodes were fabricated on the epilayer surface), the mid-depth of the wafer, and the backside of the wafer, respectively. A red pattern representing the electrode was also placed on the front surface. The definitions of the x, y, and z axes are the same as those shown in Figure 1(c).

Based on Supplementary Movie 3, it is evident from the camera view (corresponding to Figure 3(b)) that the inclination angle of the vertical grid relative to the y-direction changes as χ varies, while all vertical grids remain parallel to each other throughout the rotation. The displacement of the vertical grid among the three layers is caused by the ω rotation of 24°. At the same time, the horizontal grid (tilted 24° from the x-direction toward the +z direction) also changes with χ. When $\chi < 0°$, the horizontal grid on the front surface of the wafer appears lower in the camera view than its counterpart on the backside. At $\chi = 0°$, the horizontal grids of all three layers overlap. As χ increases further ($\chi > 0°$), the horizontal grid on the backside appears lower. This simulation clearly demonstrates that the relative position in the y-direction as χ varies provides information about the depth (z-direction) of the dislocations.

By comparing the simulated results with the experimental ones, dislocations near the front surface can be clearly distinguished from those located deeper within the sample. An example is shown in Figure 4 at nine χ angles. At $\chi = -20°$, the dislocation e1 (marked by the green arrow) appears lower than dislocation s1 (marked by the blue arrow). Their relative positions along the y-direction are reversed at $\chi = 20°$. In contrast, the relative position between e1 and e2 (marked by the red arrow) remains unchanged throughout this process, consistent with χ rotation under the assumption that the two dislocations lie on (001) planes in close proximity. In addition, the relative position between e1 (or e2) and the horizontal edges of the electrodes also remains constant. Since the electrodes are located on the epilayer surface, this indicates that e1 and e2 are dislocations in the epilayer, whereas s1 is a dislocation in the substrate. A complete high-resolution dataset is provided in Supplementary Movie 4 and Supplementary Figure 1.

With the ability to distinguish dislocations located at different depths, we can analyze the characteristic behavior of dislocations propagating from the substrate into the epilayer. As a general trend, very few threading dislocations, i.e., dislocations nearly perpendicular to the (001) plane, were observed. Most dislocations lie on the (001) plane, predominantly extending along the ***b***-axis, i.e., the ⟨010⟩ direction. This observation is consistent with previous reports indicating that ⟨010⟩{001} is an active slip system in β-$Ga_2O_3$ [**16**, **18**, **47**, **48**]. Since no clear penetration of dislocations from deep within the substrate into the epilayer was observed, it is likely that only those dislocations located very close to the substrate surface (i.e., near the substrate/epilayer interface) influence subsequent epitaxial growth. This suggests that greater attention should be paid to in-plane dislocations rather than threading dislocations when analyzing the impact of lattice defects on SBDs fabricated on the (001) face of β-$Ga_2O_3$. It is also worth noting that tangled dislocations in the substrate [**23**], which typically exhibit thick and dark contrast in XRT images, tend to induce the formation of complex dislocation structures in the epilayer. These structures generally appear as tangled dislocation complexes rather than isolated dislocation lines. Two such examples are shown in Figure 5. Readers are referred to Supplementary Movies 5 and 6 for animations of these examples, in which tangled epilayer dislocations (labeled “e”) can be observed near tangled substrate dislocations (labeled “s”). Due to the tangled nature of these dislocations, it is difficult to establish a clear correlation between dislocations in the substrate and those in the epilayer. However, our previous study showed that when tangled dislocations outcrop at the substrate surface, they typically behave as complexes of multiple threading dislocations [23]. This issue will be further clarified in future work.

Finally, we note that since each point $P(x, y, z)$ in the 3D crystal has a unique corresponding coordinate $P''(x'', y'', z'')$ after two sequential rotations of ω and χ, its projected image on the camera therefore has a unique two-dimensional coordinate ($x_1$, $y_1$). This relationship can be expressed by the following equations, where $R(\omega)$ and $R(\chi)$ are the rotation matrices corresponding to ω and χ, respectively:

$$P''(x'', y'', z'') = R(\omega)R(\chi)P = \begin{pmatrix} \cos\omega & \sin\omega\sin\chi & \sin\omega\cos\chi \\ 0 & \cos\chi & -\sin\chi \\ -\sin\omega & \cos\omega\sin\chi & \cos\omega\cos\chi \end{pmatrix} \begin{pmatrix} x \\ y \\ z \end{pmatrix} \quad \text{(Eq. 1)}$$

$$x_1 = \cos\omega \cdot x + \sin\omega\sin\chi \cdot y + \sin\omega\cos\chi \cdot z \quad \text{(Eq. 2)}$$

$$y_1 = \cos\chi \cdot y - \sin\chi \cdot z \quad \text{(Eq. 3)}$$

From Eqs. 2 and 3, it is evident that ($x_1$, $y_1$) depends on χ and $z$ for a given ($x$, $y$). Therefore, by acquiring a series of χ-dependent topo-tomographic images, depth information along the $z$-direction can be extracted. The datasets in this study further enable full 3D reconstruction of dislocations in β-$Ga_2O_3$ using algorithms developed for cone-beam X-ray topo-tomography and X-ray laminography [39, 40, 49, 50]. Details of the reconstruction method, based on the inverse Radon transform [38] and the Fourier slice theorem [51], will be reported elsewhere.

## IV. CONCLUSIONS

In summary, we demonstrated 3D visualization of dislocations in β-$Ga_2O_3$ using synchrotron-radiation X-ray topo-tomography under a two-beam Borrmann-effect condition in transmission XRT. By rotating the sample about the diffraction vector and analyzing the evolution of dislocation contrast, depth-resolved information on dislocation distributions was obtained. This approach enabled clear distinction between dislocations located in the epilayer and those in the substrate of SBD structures. The results show that most dislocations lie on the (001) plane and predominantly extend along the ⟨010⟩ direction, while threading dislocations are relatively scarce. In addition, no clear propagation of dislocations from deep within the substrate into the epilayer was observed, suggesting that dislocations near the substrate/epilayer interface play a dominant role in subsequent epitaxial growth.

Furthermore, tangled dislocations in the substrate were found to induce complex dislocation structures in the epilayer. These findings highlight the importance of controlling near-interface and in-plane dislocations for improving $\beta$-$Ga_2O_3$-based device performance.

The present method provides an intuitive and practical approach for depth-resolved characterization of lattice defects and offers a useful tool for understanding defect behavior in $\beta$-$Ga_2O_3$ and other crystalline materials.

**Supplementary Material**

See Supplementary Movies 1 and 2 for the geometrical simulation results of dislocation set 1 (spindle-shaped dislocation) and set 2 (three dislocations No. 1–3 with varying angles relative to [100]).

See Supplementary Movie 3 for an animation of three datasets—the 3D view, the camera view of the simulated results, and the experimentally acquired topo-tomographic images—over the range of $\chi = -20°$ to $+20°$, with $\chi$ increments of 0.5° per step.

See Supplementary Movie 4 and Supplementary Figure 1 for a complete high-resolution dataset of 81 topo-tomographic images recorded at 0.5° intervals over the $\chi$ range from −20° to +20°.

See Supplementary Movies 5 and 6 for animations of X-ray topo-tomographic images obtained from two representative areas, in which tangled epilayer dislocations are observed near tangled substrate dislocations.

**Acknowledgments**

This study was financially supported by (1) New Energy and Industrial Technology Development Organization (NEDO) project, No. JPNP22007; (2) JSPS KAKENHI Grant No. 20K05355, 23H01872, and 23K17356 Japan, (3) the Sumitomo Foundation, Japan, (4) the Kurata Grants by the Hitachi Global Foundation, (5) the Iketani Science and Technology Foundation, and (6) The Iwatani Naoji Foundation. The authors thank colleagues at NCT for preparing the samples. The synchrotron XRT observations were performed at KEK-PF under proposal Nos. 2020G585, 2022G503, and 2024G520. Y.Y. gratefully acknowledges Prof. Dr. K. Hirano for his support in the XRT experiments.

## AUTHOR DECLARATIONS

### Declaration of competing interests

The authors declare that they have no known competing financial interests or personal relationships that could have appeared to influence the work reported in this paper.

## Generative AI

Not used in the manuscript preparation process.

## Author contributions

Yongzhao Yao: Conceptualization (lead); Data curation (lead); Investigation (lead); Resources (equal); Writing - original draft (lead); Writing - review & editing (equal);

Daiki Katsube: Data curation (equal); Investigation (equal); Resources (equal); Writing - review & editing (equal);

Hirotaka Yamaguchi: Data curation (equal); Investigation (equal); Resources (equal); Software (lead); Writing - review & editing (equal);

Shinya Yamaguchi: Resources (equal); Writing - review & editing (equal);

Daiki Wakimoto: Resources (equal); Writing - review & editing (equal);

Hironobu Miyamoto: Resources (equal); Writing - review & editing (equal);

Yukari Ishikawa: Funding acquisition (lead); Project administration (lead); Resources (lead); Writing - review & editing (equal);

All authors have approved the manuscript.

## Data Availability

Raw data were generated at the synchrotron facilities KEK-PF. The data that support the findings of this study are available within the article and its supplementary material.

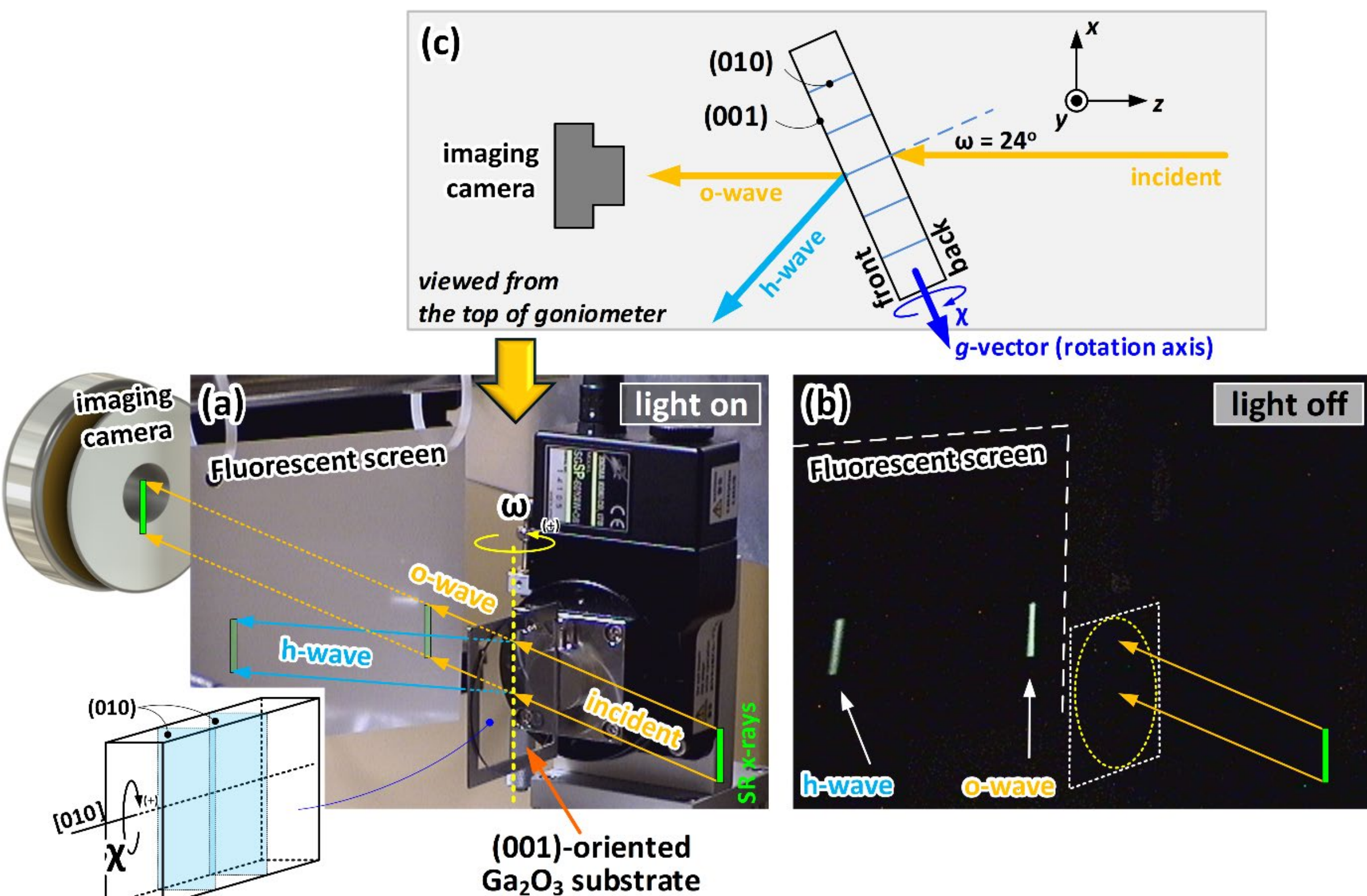

**Figure 1.** (a) Photograph of the optical system; (b) Photograph of the fluorescent screen under the Bragg condition for the Borrmann effect; (c) Geometrical configuration of the system viewed from the top of the goniometer.

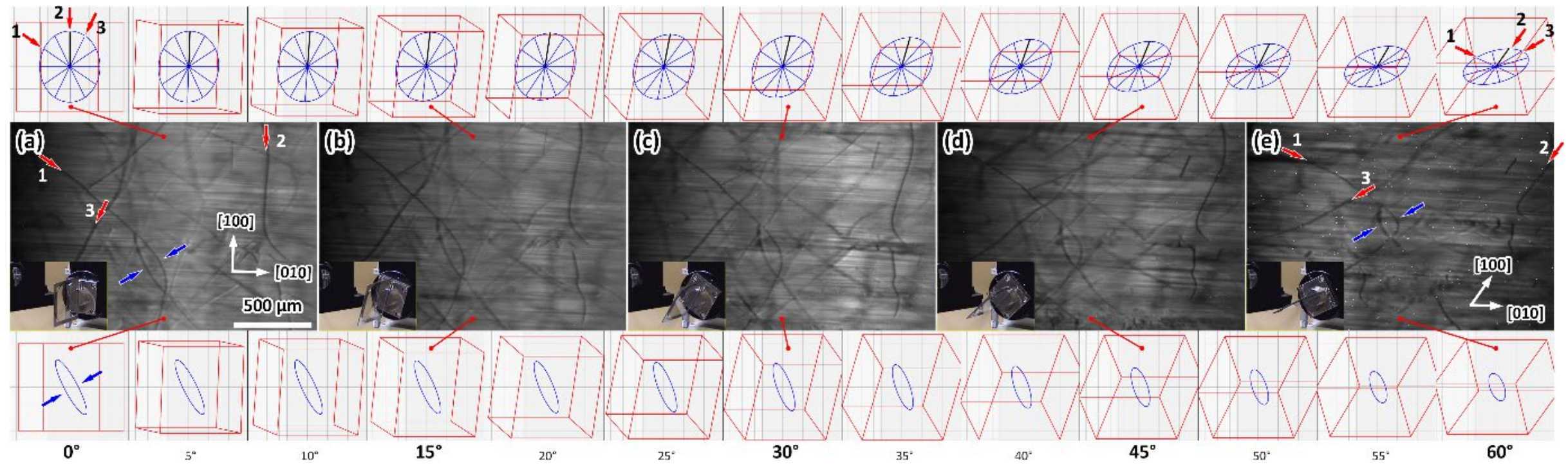

**Figure 2.** (a)–(e) X-ray topo-tomography images acquired with χ rotated in 15° steps from 0° to 60°. All images share a 500 μm scale bar. Insets in each panel show photographs of the sample. The graphs above and below the images present the corresponding simulated results for varying χ. The spindle-shaped dislocations indicated by two blue arrows and dislocations No. 1–3, marked by red arrows, show good agreement between experiment and simulation, suggesting that they lie on the (001) plane.

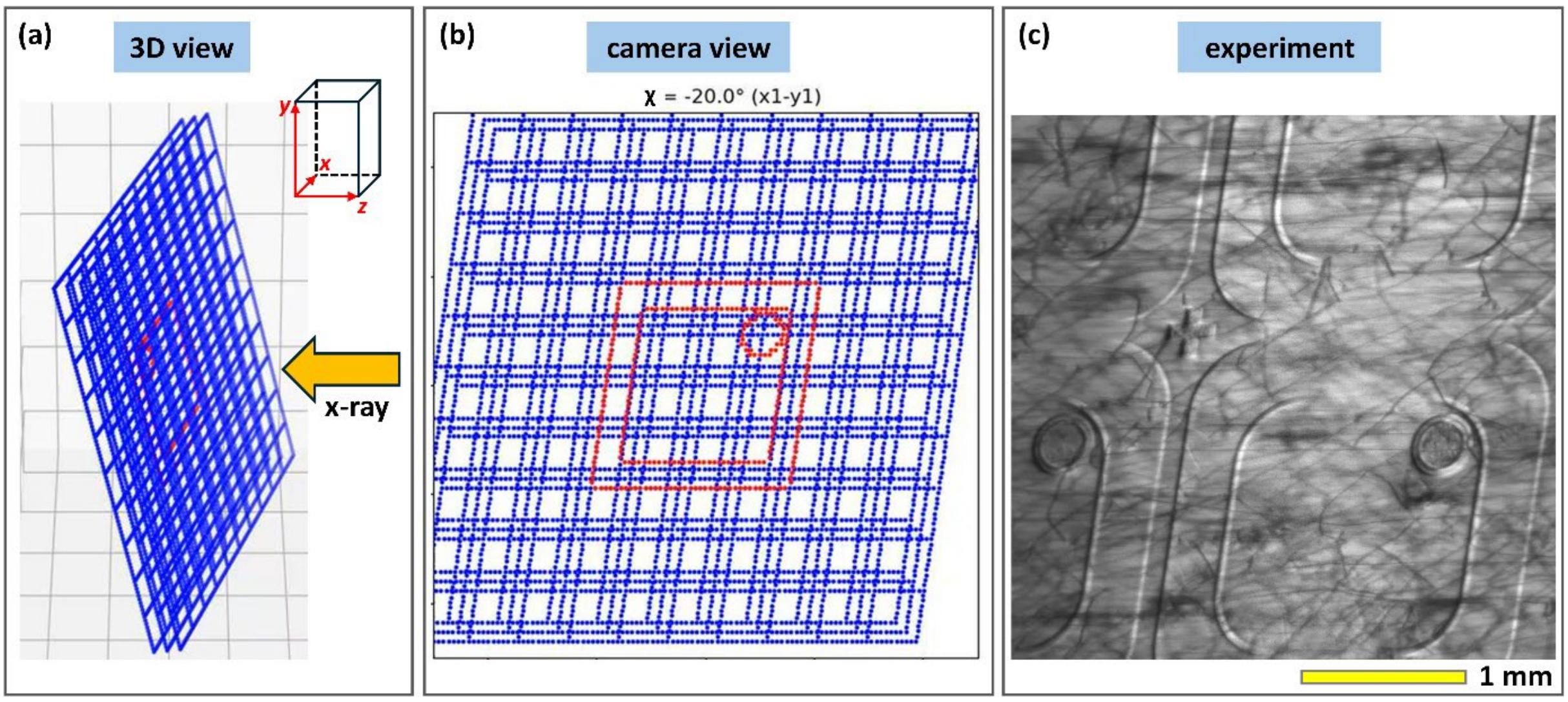

**Figure 3.** (a) 3D view of the simulated model, (b) corresponding camera view, and (c) experimentally acquired X-ray topo-tomographic image at $\chi = -20°$. See Supplementary Movie 3 for the full animation over $\chi = -20°$ to $+20°$.

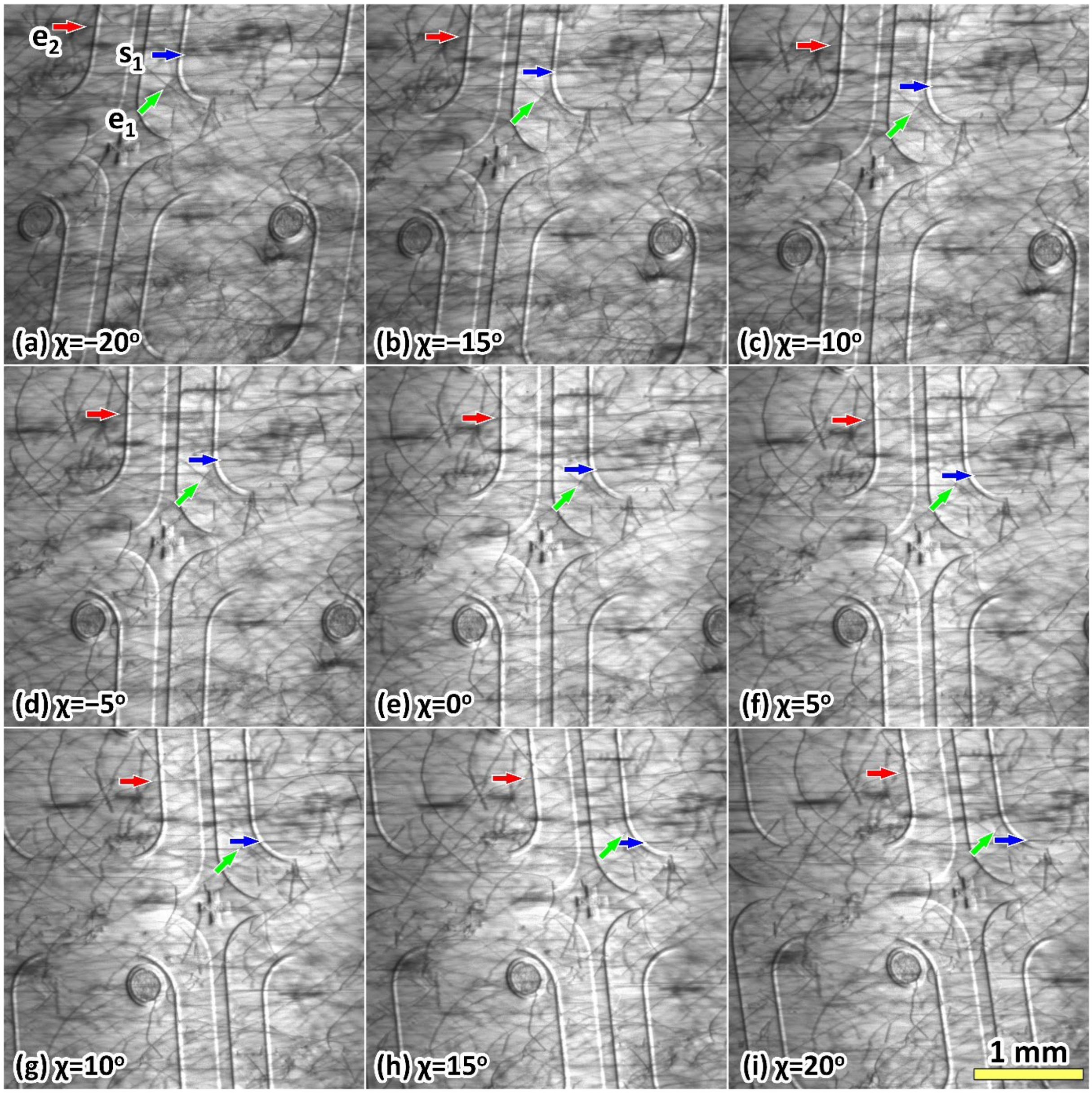

**Figure 4.** Topo-tomographic images at nine different χ angles. Dislocations e1 and e2 (epilayer, marked by green and red arrows) and s1 (substrate, marked by a blue arrow) exhibit different χ-dependent positional shifts along the y-direction. See Supplementary Movie 4 for the full animation over χ = −20° to +20°.

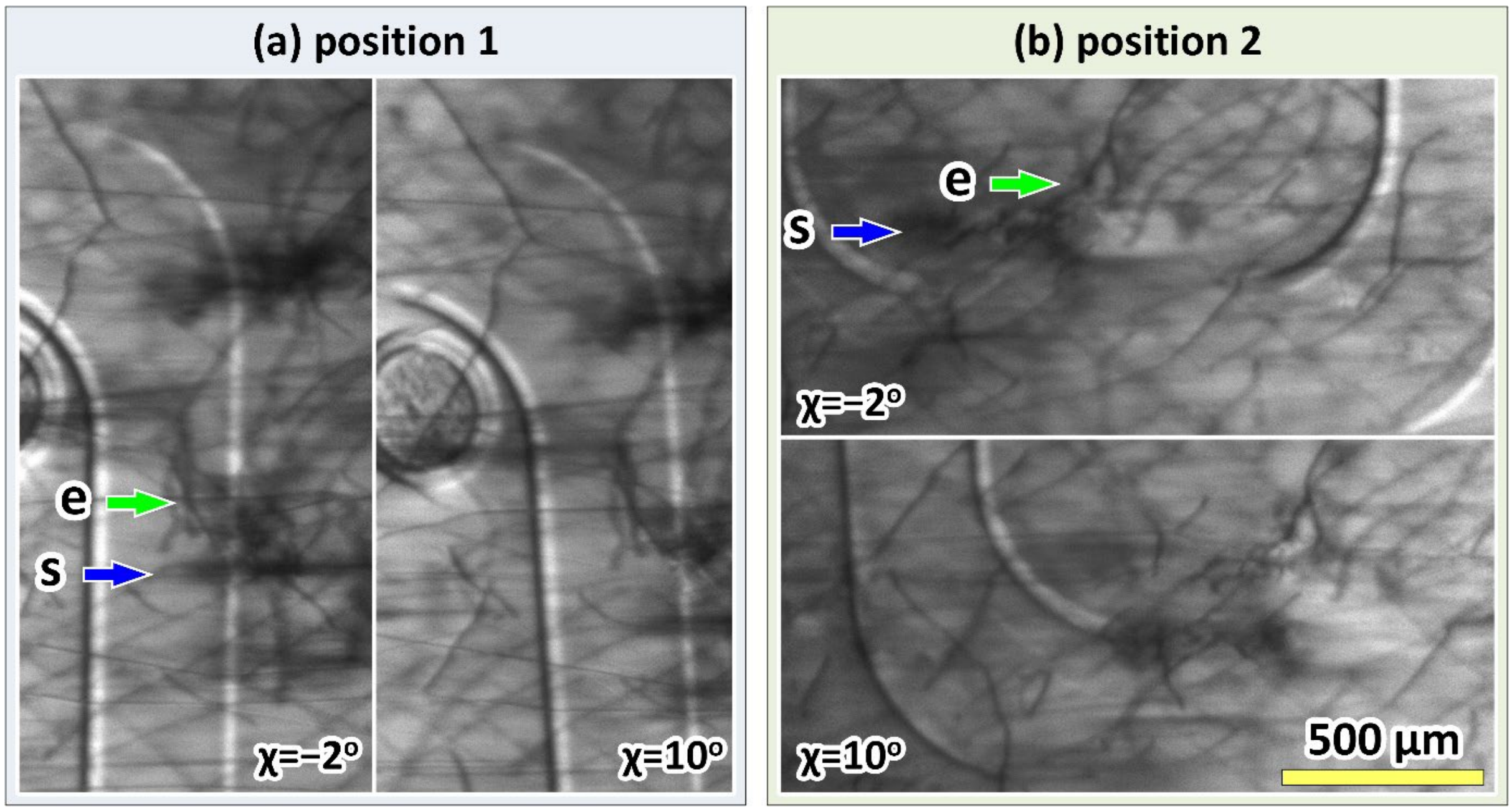

**Figure 5.** Examples of tangled dislocations observed in both the substrate and epilayer. Tangled epilayer dislocations (labeled "e") are observed in close proximity to tangled substrate dislocations (labeled "s"), suggesting a correlation between complex dislocation structures across the substrate/epilayer interface. See Supplementary Movies 5 and 6 for animations of the two representative areas.